\documentclass{PoS}
\usepackage{amsfonts,amssymb,amsmath,amssymb}

\title{Unravelling the non-standard top and Higgs couplings in associated top-Higgs production at the High-luminosity LHC}

\ShortTitle{Top Asymmetries and anomalous couplings}

\author{Saurabh D. Rindani\\
        Theoretical Physics Division, Physical Research Laboratory,\\
        Navrangpura, Ahmedabad 380 009, India,\\
        E-mail: \email{saurabh@prl.res.in}}

\author{\speaker{Pankaj Sharma}\\
        Center of Excellence in Particle Physics (CoEPP),\\
        The University of Adelaide, Adelaide, Australia\\
        E-mail: \email{pankaj.sharma@adelaide.edu.au}}
        
\author{Ambresh Shivaji\\
INFN, Sezione di Pavia,\\ Via A. Bassi 6, 27100 Pavia, Italy \\        E-mail: \email{ambresh.shivaji@pv.infn.it}}

\abstract{We study the sensitivities of non-standard top and Higgs couplings in the $pp\to thj$ process at
the 14 TeV high-luminosity run of Large Hadron Collider (HL-LHC). We calculate top polarization
and construct various lab-frame observables to study their sensitivities on anomalous couplings at
HL-LHC. Both anomalous $Wtb$ and $hVV$ couplings contribute in the production process.
However, only the former contributes in top decay and only  the latter takes part in Higgs decay.
We study this interesting interplay of couplings in the production and decay observables.
In production, these couplings significantly affect the top polarization. As a measure of top
polarization, we look at decay-lepton angular distributions in the laboratory frame and study the
effect of anomalous couplings on these distributions. We construct certain asymmetries to study
the sensitivity of these distributions to top quark couplings. In particular, real and imaginary
parts of an $hVV$ coupling and real part of a $Wtb$ coupling $f_{2R}$ significantly alter the top
polarization and, in turn, affect the decay-lepton distributions in laboratory frame.
}

\FullConference{8th International Workshop on Top Quark Physics, TOP2015\\
		14-18 September, 2015\\
		Ischia, Italy}

\begin{document}

\section{Introduction}
To characterize the new boson discovered at the Large hadron Collider (LHC), we need to study its couplings with a 
high precision with the SM particles. In the presence of CP violation, it would have both scalar and pseudoscalar couplings. 
To determine the composition of the Higgs, in this work, we study the polarization of the top quark when it is produced in association 
with Higgs and a forward jet. This process would receive the contributions of various anomalous couplings of the Higgs and top quark.
But the most important one is top-Higgs Yukawa coupling. Direct determination of this coupling is one of the important goal of high-luminosity 
LHC (HL-LHC). 

Top polarization as a probe of this coupling in associated production of top and Higgs has been proposed in \cite{Ellis:2013yxa,Yue:2014tya}. 
In this work, we propose an azimuthal asymmetry in the laboratory frame to probe and determine the CP-phase of $t\bar th$ coupling. 
The azimuthal distribution of charged lepton in lab frame shows sensitivity to top polarization and thus to other anomalous couplings.
More details about top polarization and azimuthal asymmetry can be found in \cite{delDuca:2015gca, Godbole:2010kr} and references therein. 
It has been studied extensively in the context of constraining top chromo-dipole couplings in top pair production 
\cite{Biswal:2012dr}, in $tW$ production \cite{Rindani:2015vya}; and anomalous $Wtb$ couplings \cite{Rindani:2011pk} and 
CP violation \cite{Rindani:2011gt} in single-top production at the LHC. In the context of two Higgs doublet models (2HDM), it has been 
used to determine $\tan\beta$ \cite{Huitu:2010ad, Godbole:2011vw} and distinguish different 2HDMs \cite{Rindani:2013mqa}.

The most general $t\bar t h$ coupling is written as:
\begin{equation*}
\mathcal L_{t\bar t h} = \frac{m_t}{v}\bar t \ (\cos\zeta_t + i\sin\zeta_t)t \ h 
\end{equation*}

$\zeta_t$ is the CP-violating phase. $\zeta_t=0$ corresponds to pure scalar state while $\zeta_t=\pi/2$ to pure pseudoscalar state.
Current LHC data on $h\to \gamma\gamma$ and $h\to g g$ constrains $|\zeta_t| <0.6\pi$

Using only Lorentz and gauge invariance, the $WWh$ couplings upto dimension-5 can be written as:
\begin{eqnarray*}
\mathcal L_{WWh}&=&g_{h1}(G^+_{\mu\nu}W^{-\mu}+G^-_{\mu\nu}W^{+\mu})\partial^\nu h+g_{h2}(G^-_{\mu\nu}G^{+\mu\nu})h \\
&-& g_{h3}\frac{m_W^2}{v}(W^+_\mu W^{-\mu})h + g_{h4}(G^+_{\mu\nu}\tilde{G}^{-\mu\nu} - G^-_{\mu\nu}\tilde{G}^{+\mu\nu})h 
\end{eqnarray*}
where $G^{\mu\nu}$ and $\tilde{G}^{\mu\nu}$ are the field strength tensor and its dual respectively.
Using only Lorentz invariance, the $tbW$ coupling can be written as:

\begin{equation*}
 \Gamma^\mu =\frac{-g}{\sqrt{2}}V_{tb} \left[\gamma^{\mu}
( \mathrm{f_{1L} P_{L}+ {f_{1R}} P_{R})}-
\frac{i \sigma^{\mu \nu}}{m_W}(p_t -p_b)_{\nu}(\mathrm{f_{2L} P_{L}+f_{2R}}
P_{R})\right] \label{anomaloustbW2}
\end{equation*}
These couplings were studied in the context of associated-single top and Higgs production process in ref. \cite{Agrawal:2012ga}.

\section{Top polarization and angular asymmetries of charged lepton}
Top polarization can be determined through the angular distribution of its decay products. 
In this work, we study the decay-lepton azimuthal distribution, in the lab frame, as a qualitative measure of top polarization. 
In the lab frame, we define the lepton azimuthal angle with respect to the top-production plane chosen as the $x$-$z$ plane,
with the convention that the $x$ component of the top momentum is positive. 

\begin{figure}[h!]
\begin{center}
\includegraphics[width=\textwidth]{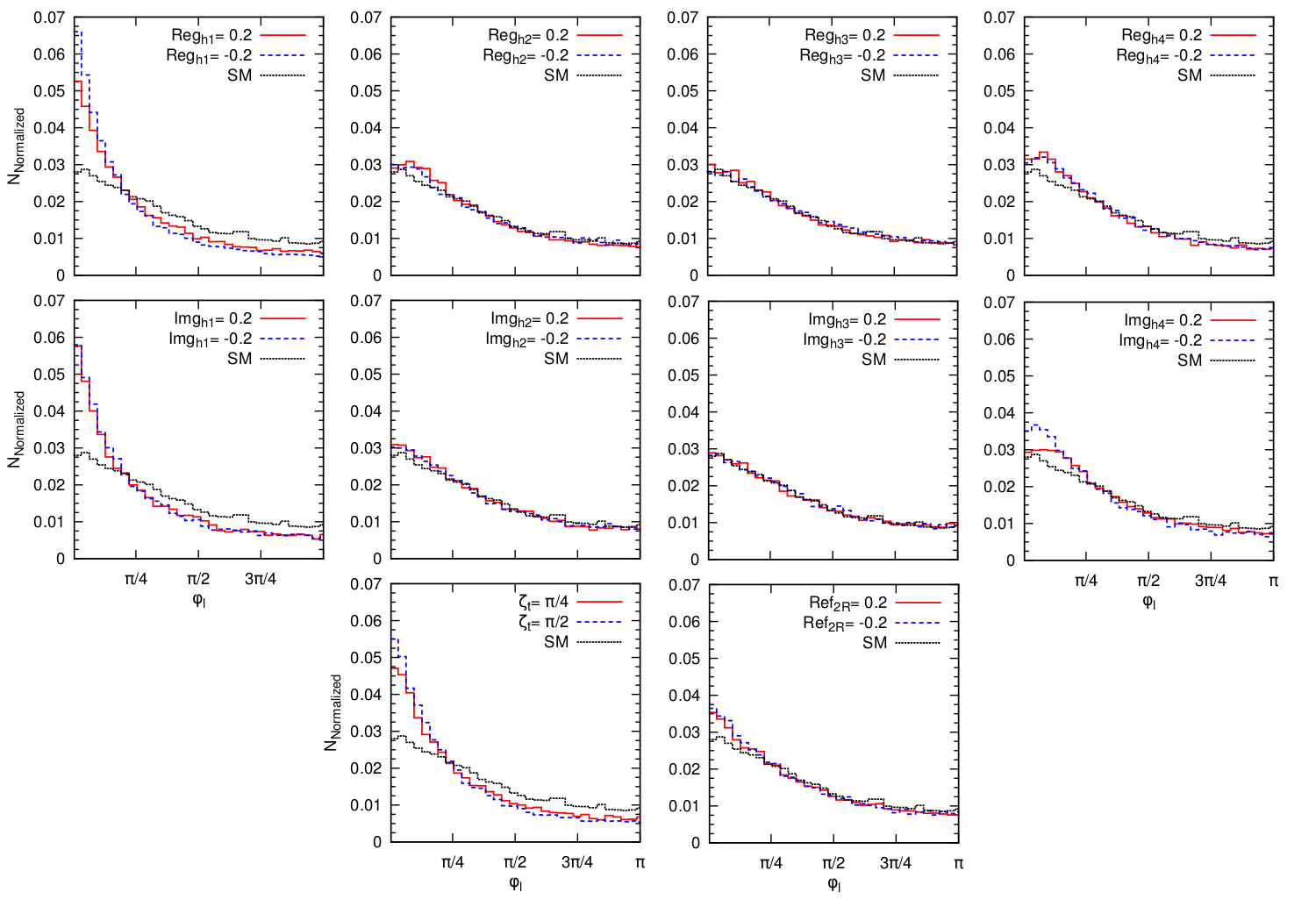}\vspace{-0.75cm}
\caption{ The normalized azimuthal distribution of the charged lepton in $thj$
production at the LHC14 for anomalous $WWh$, $Wtb$ and $tth$ couplings. Also shown in each case is the distribution for the SM.} 
\label{dist-phi}
\end{center}
\end{figure}

In Fig. \ref{dist-phi}, we show the lepton azimuthal distribution for different anomalous top couplings at the 14 TeV LHC. 
We find that this distribution is sensitive to CP-phase of top Yukawa coupling; and real and imaginary parts of the coupling $g_{h1}$.
To quantify the sensitivity of these  
distributions, we further define an asymmetry in terms of partially integrated cross sections
\begin{equation}
 A_{\phi}^\ell=\frac{\sigma(\cos \phi_\ell >0)-\sigma(\cos
\phi_\ell<0)}{\sigma(\cos \phi_\ell >0)+\sigma(\cos \phi_\ell<0)},
\label{aziasy}
\end{equation}

In Fig. \ref{aziasy}, we display the top polarization and azimuthal asymmetry for different anomalous top couplings 
at the 14 TeV LHC. One can easily notice from the top and bottom panel of fig. \ref{aziasy} the fact that the asymmetry, $A_{\phi}^\ell$, 
reconstructs the qualitative behaviour of the top quark polarization. The $A_{\phi}^\ell$ is quite sensitive to real and imaginary parts of 
$gh_1$ and CP-phase of top-Yukawa coupling as can be seen from the fig. \ref{aziasy}. After the HL-LHC run, the total integrated luminosity 
is expected to be about 3 ab$^{-1}$. With that amount of data, this asymmetry would determine the CP phase to within $\pi/16$. 

\begin{figure}[h!]
\begin{center}
\includegraphics[width=0.32\textwidth]{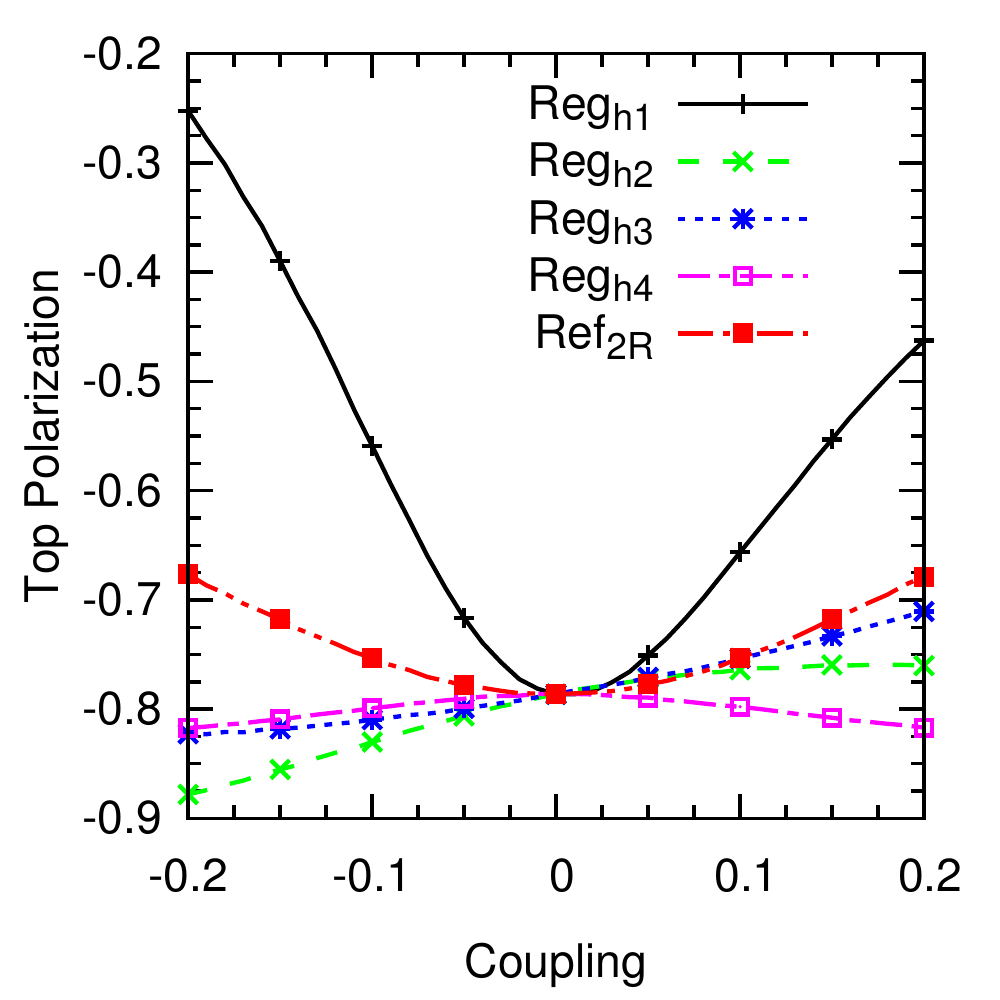} 
\includegraphics[width=0.32\textwidth]{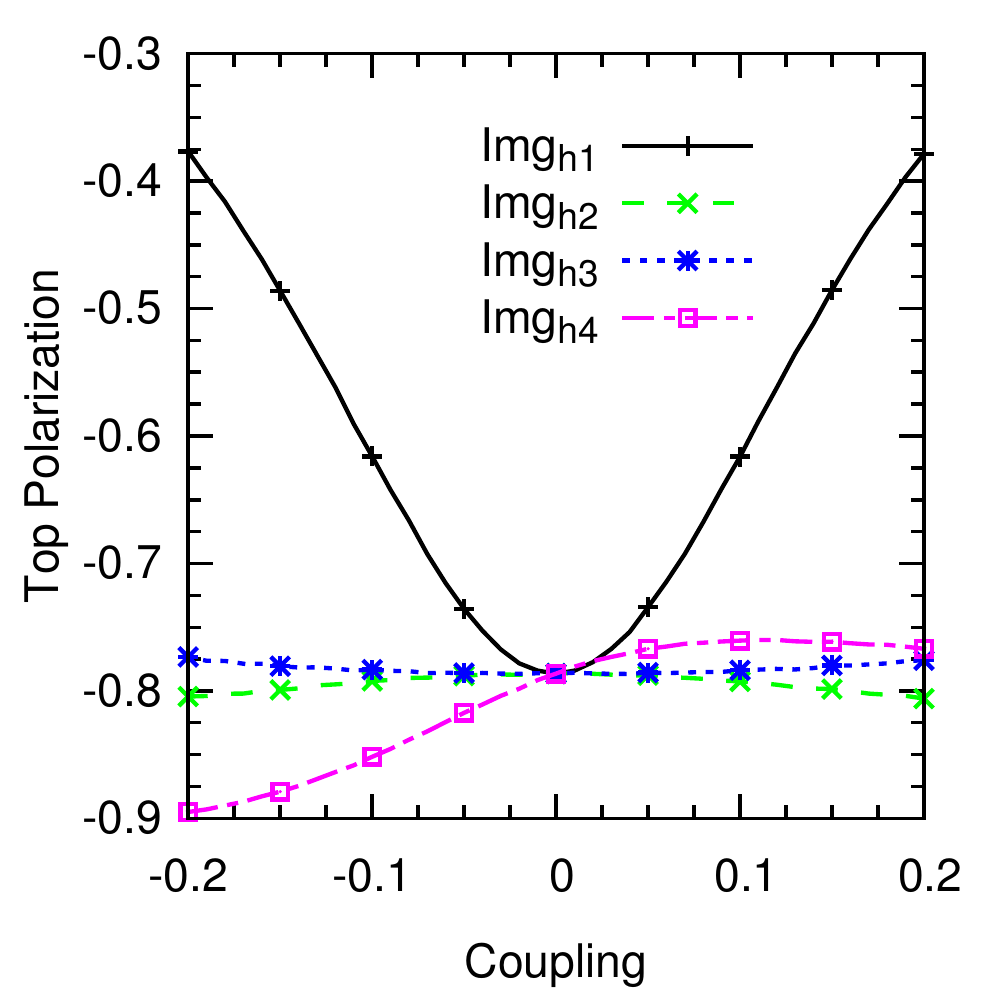}
\includegraphics[width=0.32\textwidth]{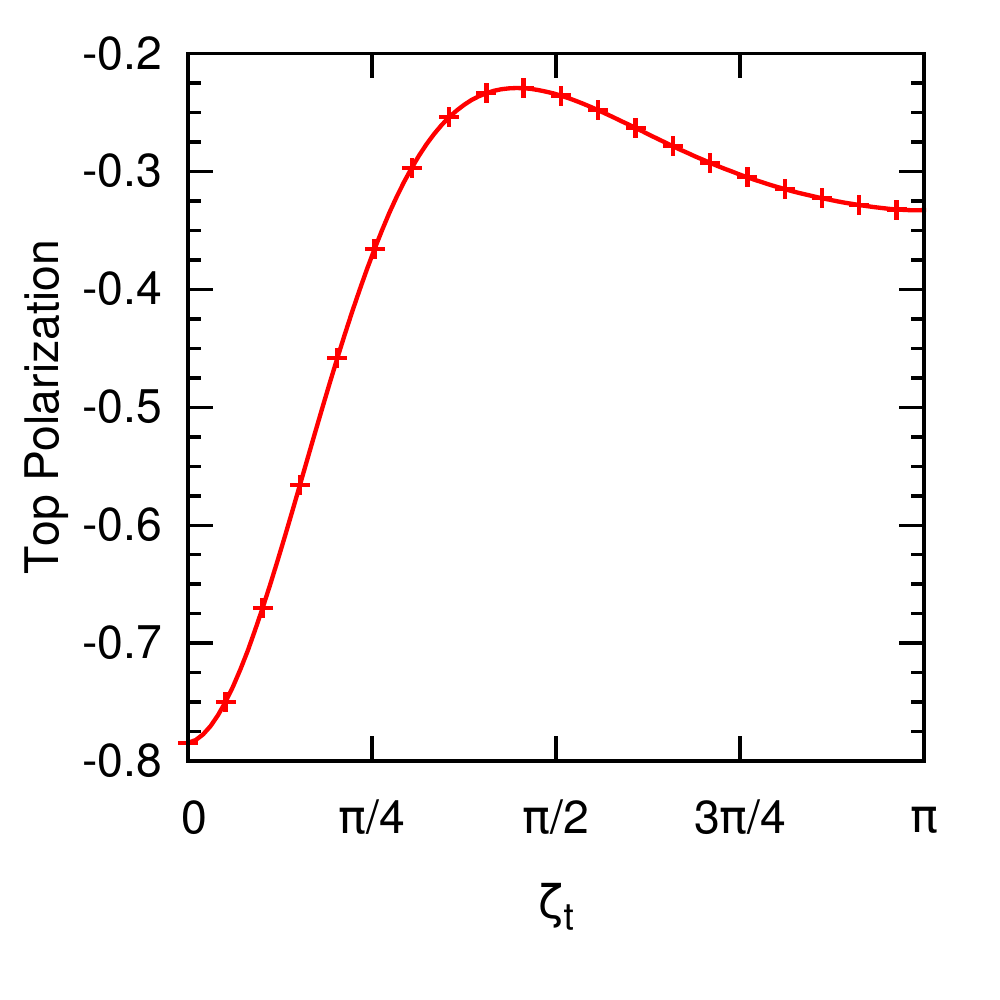}\\
\includegraphics[width=0.32\textwidth]{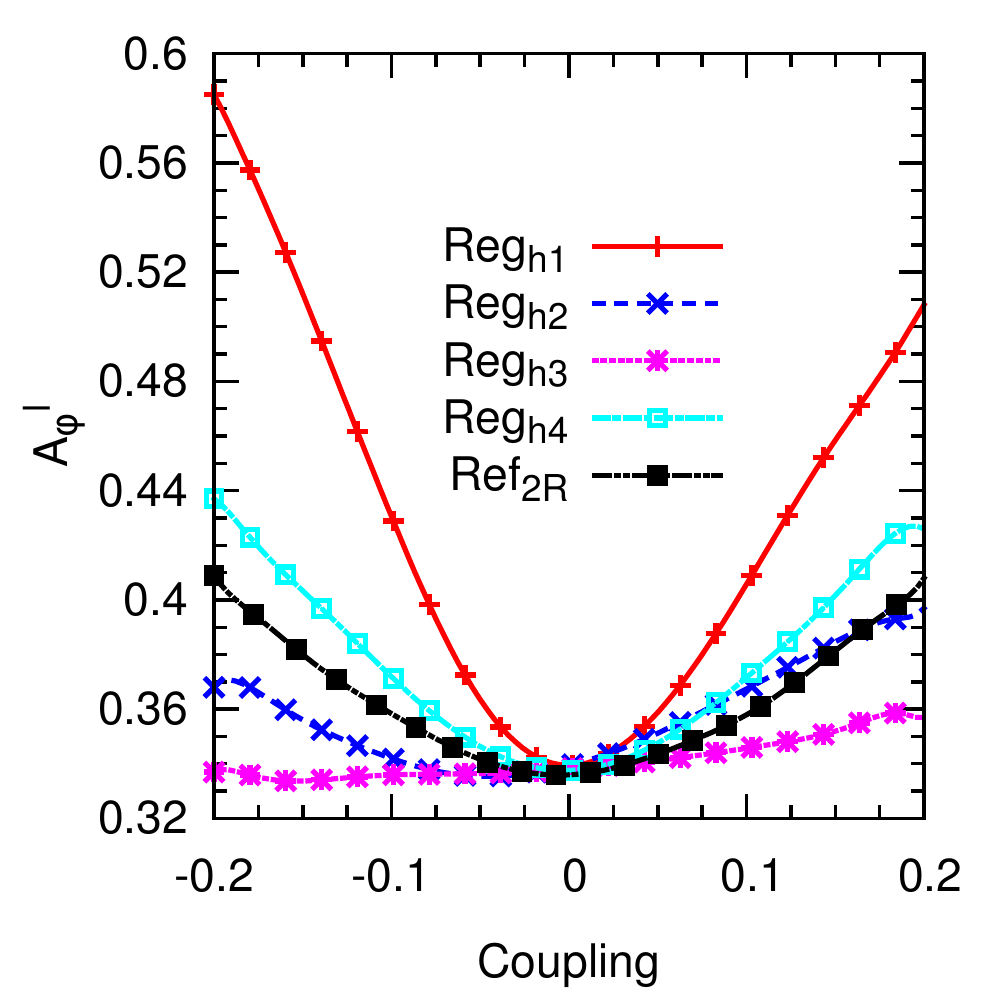} 
\includegraphics[width=0.32\textwidth]{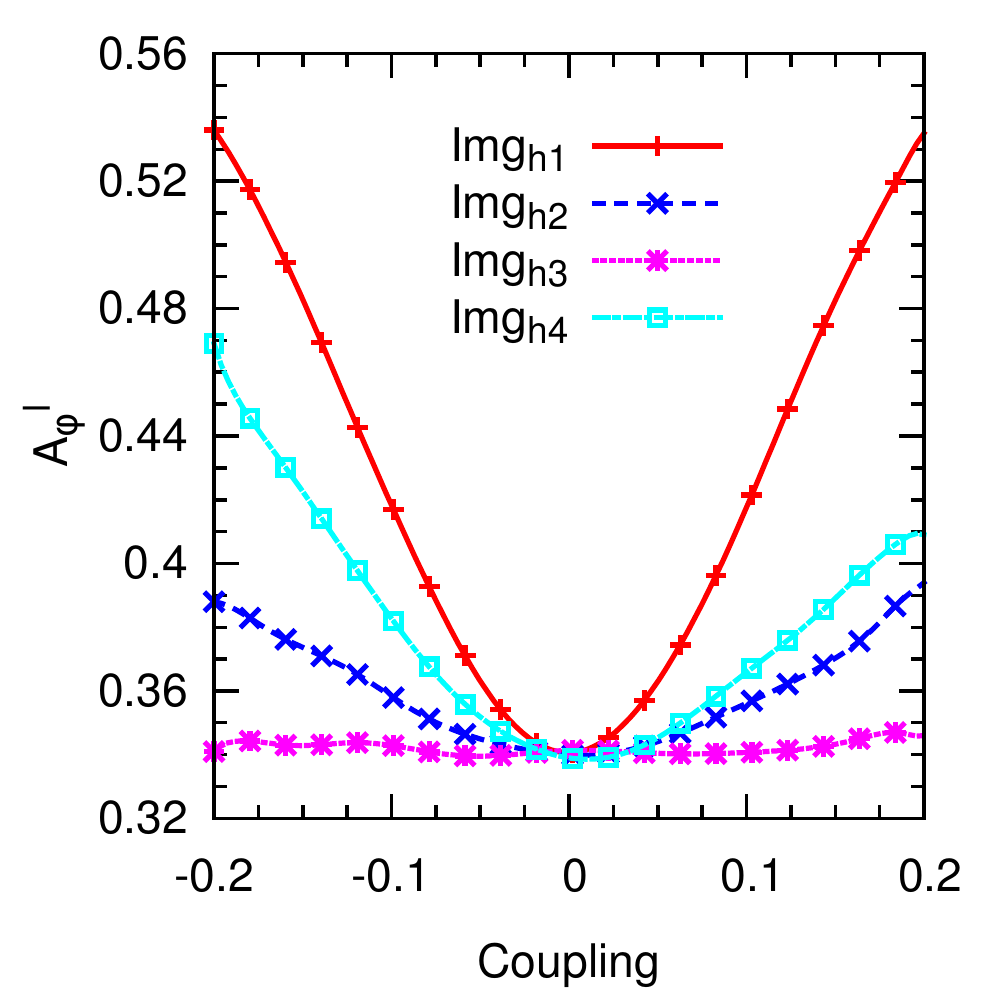}
\includegraphics[width=0.32\textwidth]{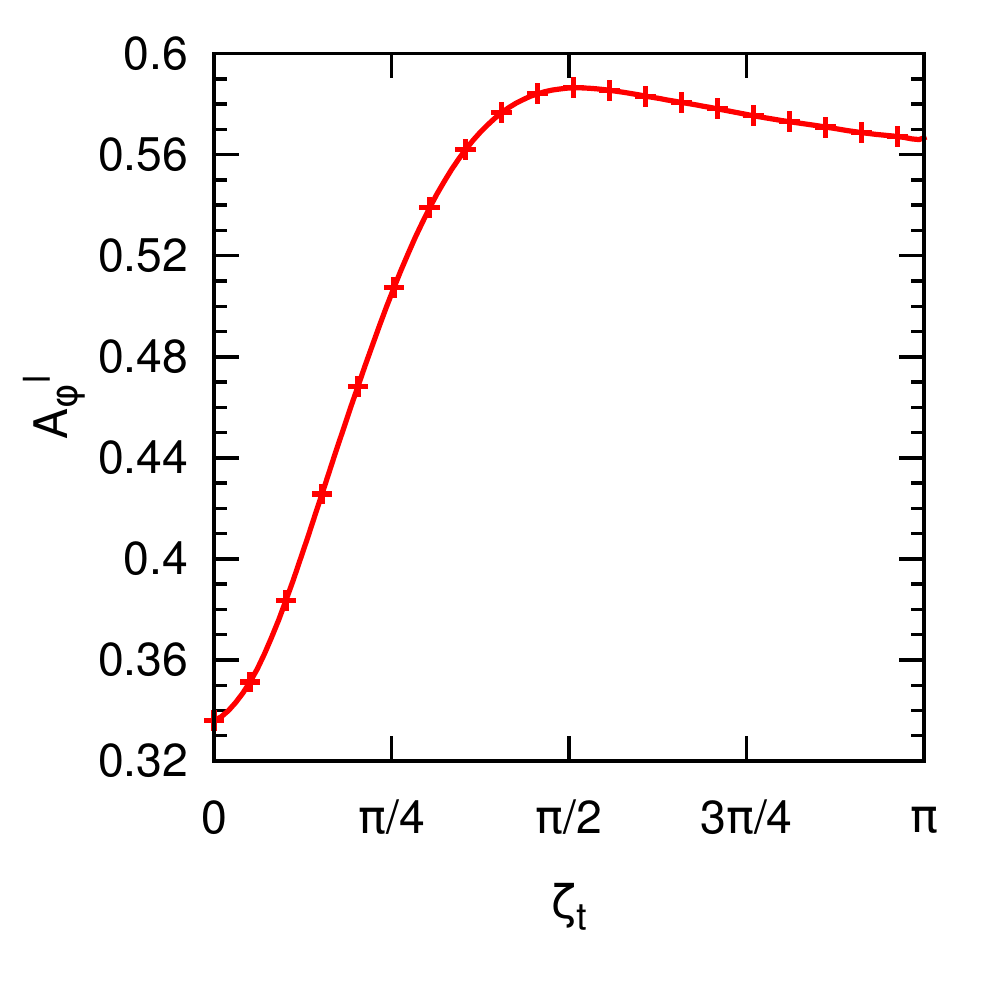}
\end{center}\vspace{-0.75cm}
\caption{ The top polarization and azimuthal asymmetry $A_\phi$ for $pp\to thj$ production as a function of top- and Higgs-anomalous couplings. } 
\label{aziasy}
\end{figure}

\section{Conclusions}\label{conclusions}
In conclusion, we find that the azimuthal asymmetry, which is defined in the lab frame, would be a better probe to determine the CP-structure of the
newly-discovered boson than the rest-frame asymmetries. The top polarization and the azimuthal asymmetry is sensitive to 
CP phase in the parameter space where cross section is not much sensitive.

\end{document}